\documentclass[preprint,aps,showpacs]{revtex4}
\usepackage{graphicx}
\usepackage{amsmath}
\usepackage{bm}
\begin{document}

\title {Atomic Theory of the Two-fluid Model: Broken Gauge
Symmetry in Bose-Einstein condensation}

\author{S. J. Han}

\affiliation{P.O. Box 4671, Los Alamos, NM 87544-4671}

\begin{abstract}

We discuss the collective excitations in a spatially inhomogeneous
(cylindrically symmetric) Bose-Einstein condensation (BEC) at low
temperature ($T \ll T_{\lambda}$). The main result is the
dispersion relation for a (first) sound wave that is obtained by
describing the perturbation as a Lagrangian coordinate. The
dispersion curve is in good agreement with the Bogoliubov phonon
spectrum $\omega=c\,k$, where $k=k_{\theta}=m/r$, the wave number
and $c=[4 \pi a \rho \hbar^{2}]^{1/2}/M$, the speed of first
sound. Based on Bohm's quantum theory, a spontaneously broken
gauge symmetry in a quantum fluid is discussed in terms of the
quantum fluctuation-dissipation, from which it is shown that the
symmetry breaking takes place at the free surface of BEC in an
external field.

\end{abstract}

\maketitle

\section{\label{sec:level1} Introduction}

The Bose-Einstein condensation (BEC) is a remarkable quantum
phenomena occurring in a macroscopic Bose system \cite{Lieb02}.
London has proposed that the $\lambda$ transition between He I and
He II is a result of the same process which causes the
condensation of an ideal Bose gas at low temperatures
\cite{London38}. In 1941, Landau studied the collective excitation
spectrum in He II based on his phenomenological theory of
two-fluid model \cite{Landau41}, and has shown that the excitation
spectrum $\omega(k)$ of a wave number $k$ should rise linearly
with slope $c$ [{\it i.e.,} $\omega=ck$ with the speed of (first)
sound $c=238 m/sec$] for small $k$, pass through a maximum, drop
to a local minimum at some value $k_{0}$, and then rise again for
$k>k_{0}$, which has been confirmed in great detail by Henshaw and
Woods \cite{Henshaw61}. For small $k$, the excitations are called
phonons (quantized sound wave).

In the absence of a satisfactory microscopic theory, there has
been considerable development of phenomenological theories
following Landau's two fluid model. Of the various attempts to
account for the $\lambda$ transition, the best known is that of
Feynman \cite{Feynman53}. Based on the exact quantum partition
function as an integral over particle trajectories by his
path-integral approach to quantum mechanics, Feynman studied the
two-fluid model of He II in a strongly interacting $^{4}$He gas
and pointed out that London's view on BEC is essentially correct.
The point in his argument is that, in a liquid like quantum
system, the symmetry of a Bose system plays far more important a
role than that of strong inter-atomic forces which do not prevent
these particles move freely in the system; yet the
pair-interaction which brings about $\lambda$-transition also
ensures that He II possess the collective excitations (phonons and
rotons) \cite{Feynman53}. It is also the essential point of
Bogoliubov's theory of superfluidity which has laid the basis for
much of our theoretical understanding about superfluidity of He II
\cite{Bog47,Note1}.

Penrose and Onsager \cite{Penrose56} further extended the concept
of Bose condensation to a strongly interacting Bose gas of He II
based on the properties of a ground state wave function derived
from the assumption that there is no long-range order interaction.
They proposed that one could essentially define the Bose condensed
state of a system such as He II as a state in which the reduced
density matrix of the system can be factorized in a certain
special way, now known as the off-diagonal long range order
(ODLRO) after Yang \cite{Yang62}.

Although the previous theories \cite{Landau41,Feynman53,Bog47}
have been very successful in explaining the collective phenomena
in He II, the dispersion relation for a longitudinal (first) sound
wave $\omega=ck$ is given only in the case of {\it a spatially
homogeneous system} for mathematical convenience \cite{Note1}.
Bogoliubov \cite{Bog47} has used an ingenious mathematical method
based directly on quantum field theory for his calculation of the
excitation spectrum in a weakly interacting Bose system.

The formal extension of the Bogoliubov method to a finite
inhomogeneous system is difficult, since one can quantize a scalar
field of quasi-particles only in a Hilbert space \cite{Note1}.
Furthermore, it is impossible to make use of the elegant
mathematical technique developed in his quantum field theory
approach, because the theory of collective excitations in a finite
space problem reduces to an initial-boundary value problem for
which one must find a correct boundary condition in terms of a
dynamical variable. It is, therefore, important to use an
appropriate mathematical technique to develop a theory of
collective excitations in BEC in a trap.

Recently, however, the study of a new BEC in a trap has helped to
revive interest in the study of collective excitations, but it has
faced an almost insurmountable challenge to derive a correct
dispersion relation for collective excitations in BEC in a trap
\cite{Fetter01}. The problem with the boundary conditions
appearing in the dynamical equations represents the major
stumbling block in obtaining a proper dispersion relation for
phonon spectrum at low temperature ($T \ll T_{\lambda}$)
\cite{Note2}.

With proper boundary conditions a theory must be capable of
computing the excitation spectrum for phonons \cite{Bog47}, and
yet explains in a natural way why a symmetry braking takes place
and how the sound wave dissipates by the fluctuation-dissipation
process at the free surface. These are uniquely new phenomena in a
Bose system that is confined by an external potential. Such a
theory is the subject of this paper.

In this paper we consider an imperfect Bose gas of N identical
particles in a trap. Between each pair of any particles in the
system, there is a hard sphere repulsion of range $a$ that
balances the external potential which determines the size of the
condensation in the trap. This pair-interaction is the basic, and
almost the only, assumption to describe the weakly interacting
Bose particles in a trap. Moreover, this assumption is reasonable
to give results which are satisfactory in a dilute Bose gas in
trap \cite{Fetter01}; the results are also consistent with the
theory of superfluidity \cite{Bog47}. The main contributions of
the present paper are a semi-classical calculation of the
excitation spectrum for phonons and a study of the symmetry
breaking along with the fluctuation-dissipation process of sound
waves to conserve the energy in an isolated system. The study of
collective excitations is essential for a complete understanding
of certain properties of a Bose condensed system, in particular,
the longitudinal collective excitations (first sound) at
temperature near absolute zero. This provides the course of
experiments that confirm BEC in a trap \cite{Penrose56}.

Because of a finite, inhomogeneous density of BEC, the collective
excitation spectrum depends on the spatial density profile and the
boundary conditions. In this paper we shall study the existence
and excitation energy spectrum of a cylindrical (first) sound wave
in a cylindrically symmetric inhomogeneous Bose condensate (CSIBC)
which is an extension of the previous study \cite{SJHan1}. The
present analysis, in principle, can be extended to an ellipsoidal
Bose condensate in a trap. In practice, however, the mathematical
analysis of the collective excitations in an ellipsoidal
Bose-Einstein condensation in prolate spheroidal coordinates is
mathematically more challenging due to the complications of the
ellipsoidal wave propagations, and the well-known special cases
may also play an important role to check the general case
\cite{Dyson68,Lamb45,SJHan2}.

To present this work as a self-contained paper, we first review
the concept of ODLRO in which Penrose and Onsager suggested that
the reduced density matrix of a Bose condensed state can be
factorized,
\begin{equation}
\rho(r,r^{\prime})=\psi^{\dag}(r)\psi(r)+ \gamma(|r-r^{\prime}|
\label{Odlro}.
\end{equation}
where $\gamma\rightarrow 0$ as $|r-r^{\prime}| \rightarrow
\infty$. The single particle wave function $\psi(r)$ represents
the condensed state in ODLRO and is viewed as a function of
macroscopic dynamical variables. The single particle state
function $\psi(r)$ is also interpreted as the mean value of a
quantum particle field \cite{Anderson66}.

With the hard sphere approximation, one can show that the mean
field satisfies the nonlinear Schr\"{o}dinger equation
(Gross-Pitaevskii),
\begin{equation}
i\hbar\frac{\partial\psi}{\partial t}=-\\
\frac{\hbar^{2}}{2M}\nabla^{2}\psi + [V(\bm{x})_{ext} + g_{1}|\psi|^{2}] \psi,\\
\label{Nlse}
\end{equation}
where $V_{self}=g_{1}|\psi|^{2} =4\pi\hbar^{2}a/M |\psi|^{2}$ and
$a$ is the s-wave scattering length. It is a self-consistent
Hartree equation for the Bose condensed wave function
\cite{Fetter01}. In our model of an imperfect Bose gas, the wave
function describes a near perfect gas modified as little as
possible by the presence of the pair-interactions. Basic to the
theory of superfluidity is the idea of Bogoliubov \cite{Bog47}
that this pair-interaction which brings about the Bose condensed
state also ensures that the system possess the longitudinal
collective excitations (first sound) in BEC. We must therefore use
the effective potential that includes the self-interaction term in
Schr\"{o}dinger equation, which might be thought of as a
generalization of Bohm's theory \cite{Bohm52} to the theory of
superfluidity \cite{Bog47}, since the nonlinear term
$g_{1}|\psi|^{2}$ is invariant under a $U(1)$ group transformation
\cite{BenLee73,Weinberg96}.

\section{\label{sec:level1}Basic Equations}

In order to derive the density profile for CSIBC, we write the
equations for the ensemble average energy in the usual quantum
theory \cite{Bohm52}:

 \begin{eqnarray}
 {\cal H} =\int
 \psi^{\dagger}\left(-\frac{\hbar^{2}}{2M}\nabla^{2} +
 V(\bm{x})_{ext} +\frac{g_{1}}{2}|\psi|^{2}\right)\psi d\bm{x}, \label{grounda}\\
 {\cal E}_{ave} =\int\left(\frac{\hbar^{2}}{2M}|\nabla\psi|^{2} +
 V(\bm {x})_{ext}|\psi|^{2}+\frac{g_{1}}{2}|\psi|^{4}\right) d\bm{x},
 \int\psi^{\dagger}\psi d\bm{x}=N.\label{groundb}
 \end{eqnarray}

For a BEC in a trap, the ground state density can be obtained in
terms of an external potential and the chemical potential by
minimizing the energy functional Eq.~(\ref{groundb}) for a fixed
number of particles in the system with the condition $\bm{p}=0$
(Penrose-Onsager criterion for BEC):
\begin{equation}
\rho(\bm{x})=|\psi(\bm{x})|^{2}=\frac{M}{4\pi\hbar^{2}a}[\mu
-V(\bm{x})_{ext}], \label{Grho}
\end{equation}
where $\mu$ is a Lagrangian multiplier and is the chemical
potential. It should be noted that the ground-state wave function
Eq.~(\ref{Grho}) has a nodal surface at which the density becomes
zero and is the boundary of CSIBC.

As we have emphasized above, the most serious question in regard
to a possible application of Bogoliubbov's quantum field theory
technique (or an extended version, the Bogoliubov-de Gennes
equation) \cite{Bog47} to a finite Bose system is that the theory
remains valid only in an infinite uniform system or a system in a
box with periodic boundary conditions but not a finite
inhomogeneous system of BEC in a trap \cite{Note1}.

Both for this reason, and because it is essential to show that a
symmetry breaking takes place at the nodal surface of CSIBC, we
employ Bohm's quantum theory with emphasis on the ensemble of
particle trajectories \cite{Bohm52,Aharonov63}. Briefly, Bohm
writes $\psi(r)$ in the form
$\psi(x,t)=f(\bm{x},t)exp[\frac{i}{\hbar}S(x,t)]=
\rho^{1/2}(x,t)exp[\frac{i}{\hbar}S(x,t)]$, where $S(x,t)$ is a
phase (or an action) \cite{Bohm52,Aharonov63}. We then rewrite
Eq.~(\ref{Nlse}) to obtain,
\begin{subequations}
\label{allequations} \label{Mean}
\begin{equation}
\frac{\partial\rho}{\partial t}+\bm{\nabla}\cdot(\rho\frac{\bm\nabla S}{M})=0\\
\label{Cont0}
\end{equation}
\begin{eqnarray}
 \frac{\partial S}{\partial t} + \frac{(\bm\nabla S)^{2}}{2M} +
V(\bm{x}) - \frac{\hbar^{2}}{4M}[\frac{\nabla^{2}\rho}{\rho}-
\frac{1}{2}\frac{(\nabla\rho)^{2}}{\rho^{2}}]=0, \label{Qhje}
\end{eqnarray}
\end{subequations}
where $ V(\bm{x})=V(\bm{x})_{ext}+g\rho$. Here the potential
$V(\bm{x})$ includes $V_{self}=g\rho$. The last term $U_{eqmp}$ of
Eq.~(\ref{Qhje}) is the effective quantum-mechanical potential
(EQMP) which plays a crucial role in the discussion of a symmetry
breaking in a finite spatially inhomogeneous Bose system.

In Eq.~(\ref{Mean}), Bohm suggested that one may regard
Eq.~(\ref{Cont0}) as the conservation of current if one identifies
$\rho$ as the probability density and
$\bm{v}=\bm{\nabla}S(\bm{x})/M$. $S(\bm{x},t)$ the solution of
Eq.~(\ref{Qhje}) is the action. However, in the limit
$\hbar\rightarrow 0$, $S(\bm{x},t)$ is a solution of the
Hamilton-Jacobi equation and becomes a phase of the wave function
$\psi(x,t)$ \cite{Bohm52,Aharonov63}. In general, the solution of
the quantum Hamilton-Jacobi equation Eq.~(\ref{Qhje}) defines an
ensemble of possible particle trajectories, which can be obtained
in principle from the Hamilton-Jacobi function, $S(\bm {x})$, by
integrating the velocity, $\bm v(\bm {x})=\bm\nabla S(\bm {x})/M$.
The equation for S implies, however, that the particle is acted
on, not only by a potential $V(\bm {x})$ but also by the effective
quantum mechanical potential $U_{eqmp}$. Moreover Eq.~(\ref{Qhje})
suggests us how we choose stable particle trajectories about which
we may linearize the equations of motion. For if, as was done in
the previous study of collective excitations in He II \cite{Bog47,
Feynman53}, a system is described as an infinite uniform medium,
$U_{eqmp}$ becomes zero. On the other hand, in a finite system the
effective quantum mechanical potential diverges at the nodal
surface,

\begin{equation}
U_{eqmp}=-\frac{\hbar^{2}}{4M}[\frac{\nabla^{2}\rho}{\rho}-
\frac{1}{2}\frac{(\nabla\rho)^{2}}{\rho^{2}}]
=-\frac{\hbar^{2}}{M}\frac{\nabla^{2}f}{f}, \label{Eqmp}
\end{equation}
where $\rho(r)$ is zero at the nodal surface.

As emphasized by Bohm \cite{Bohm52}, a particle experiences the
force from $U_{eqmp}$ and fluctuates with its momentum
$\bm{p}=\bm{\nabla}S$ and energy near the surface with the degree
of divergence $U_{eqmp}=M\omega_{0}^{2}\mu_{0}/D^{2}$ as
$D=[\mu_{0}-(1/2)M\omega^{2}_{0}r^{2}]\rightarrow 0$ near the
surface. Hence $U_{eqmp}$ breaks up the phase coherence of a sound
wave in the surface layer, the thickness of which is in the order
of mean free path.

\section{\label{sec:level1} First-Order Equations}

Now we come to the question of why we must find an alternate
approach such as Bohm's quantum theory to the quantum field theory
method by which Bogoliubov has already obtained a correct energy
spectrum of collective excitations $\omega=ck$ with
$c=[4\pi\,a\rho\hbar^{2}]^{1/2}/M$ in the long wavelength limit.
The reason is obvious: one can quantize the scalar field of
quasi-particles only in a Hilbert space \cite{Bog47,Note1}, but
not in a finite inhomogeneous system such as CSIBC. By the same
reason, an extended version of the Bogoliubov theory with the mean
field, the Bogoliubov-de Gennes equation, is not also applicable
to a finite inhomogeneous system and it would lead to erroneous
results. This observation provides a deeper appreciation of
Feynman's path-integral approach in his atomic theory of the
two-fluid model although his analysis was confined to a uniform
system \cite{Feynman53}. The basic approach in both Feynman's
two-fluid model and our present approach is that a Bose system
with self-consistent interactions can be studied by the ensemble
of particle trajectories to which Bohm's quantum theory is more
useful for describing a system with a well-defined boundary. More
importantly, by Bohm's interpretation of quantum mechanics, the
solution of the Hamilton-Jacobi equation can be separated from the
solution of the quantum Hamilton-Jacobi equation in the classical
limit ({\it i.e.}, $\hbar\rightarrow 0$), and the action of the
quantum Hamilton-Jacobi equation is used only for the measurement
process to reconcile with the uncertainty principle by statistical
ensemble of particle trajectories. This unique feature allows one
to apply perturbations to the solution of the Hamilton-Jacobi
equation, and at the same time the effective quantum mechanical
potential is useful to describing the fluctuation-dissipation
process to explain the spontaneously broken local gauge symmetry
in the system.

As in the previous work \cite{SJHan1}, the perturbation to a
particle trajectory is treated as a Lagrangian coordinate to the
semi-classical solution. This leads a set of the linearized
dynamical equations from which we derive a second-order partial
differential equation in terms of a displacement vector. We solve
this differential equation for the solution of a problem in CSIBC
with correct initial-boundary conditions.

Since the present calculation of collective excitation spectrum in
CSIBC is new, and is of some interest in itself, we present it
here in full detail. We now introduce a symmetry breaking
perturbation to the particle trajectories in CSIBC in a Lagrangian
coordinate \cite{Bernstein58,SJHan82,Weinberg67}; it is defined as
$\bm{x}=\bm{x}_{0}+\bm{\xi}(\bm{x}_{0},t)$, where
$\bm{\xi}(\bm{x}_{0},t)$ is a function of the unperturbed position
of a particle in the condensate, and remains attached to the
particle as it moves.

In addition to the particle displacements, we introduce phase
coherence that relates the Bose condensed wave function (mean
field) in ODLRO to a many-body ground-state wave function,
$\bm{\xi}\cdot\bm{\nabla}S(\bm{x}_{0},t)
=\sum_{i}\bm{\xi}_{i}\cdot\bm{\bm{\nabla}_{i}}S_{0,i}(\bm{x}_{0,i},t)$,
where $S_{0,i}(\bm{x}_{0,i},t)$ is the phase of a single Bose
particle in the system. This is a necessary and sufficient
condition for the phase coherence. It also gives a simple
interpretation of the mean field and quasi-particles (coherent
excited states) in BEC. It is therefore obvious that a single
particle excitation does not occur in our study as in Feynman's
work \cite{Feynman53}. Moreover, the collective excitations we
study in this paper are {\it phase coherent sound waves}.

The first step in developing this alternate approach is to
associate with a Bose particle having precise stable particle
trajectories which are a function of position and momentum. In
this connection, it is worthwhile to note that the use of the
Hamilton-Jacobi equation in solving for the motion of a particle
is only matter of convenience and that, in principle, we can
always solve it directly by the conventional dynamical equations
with the correct initial-boundary conditions for a finite space
problem. In order to study the modified excitations in a trap, we
find it convenient to write down the basic dynamical equations.

This is most simply done by writing $\bm
{v}(\bm{x}_{0},t)=\bm{\nabla}S(\bm{x}_{0},t)/M$, then
Eqs.~(\ref{Grho}) and (\ref{Mean}) yield the following dynamical
equations:

 the equation of motion for a single particle
\begin{equation}
M(\frac{\partial}{\partial t}\bm{v}+\bm{v}\cdot \bm{\nabla} \bm
{v})=-\bm{\nabla}\mu, \label{Motion0}
\end{equation}

 the equation of continuity,
\begin{equation}
\frac{\partial}{\partial t}\rho+\bm{\nabla}\cdot(\rho\bm{v})=0,%
\label{Cont1}
\end{equation}
where $\rho$ is henceforth interpreted as the number density,

 the equation of state,
\begin{equation}
\mu(\bm{x}, t)=\mu_{\text{loc}}[\rho(\bm{x}, t)]+V_{\text{ext}},
\label{EQS0}
\end{equation}
where $\mu_{\text{loc}}[\rho(\bm{x}, t)]$ is the chemical
potential in the local density approximation \cite{Oliva89}. These
dynamical equations allow us to show that the symmetry breaking
takes place at the nodal surface and yield a correct modified
phonon spectrum in CSIBC.

A little algebra with Eqs.~(\ref{Motion0}), (\ref{Cont1}) and
(\ref{EQS0}) gives the following linearized equations:
\begin{subequations}
\label{allequations} \label{First}
\begin{eqnarray}
S(\bm{x},t)=S(\bm{x}_{0},t) +
\bm{\xi}\cdot\bm{\nabla}_{0}S(\bm{x}_{0},t)\\
\label{Firsta} \bm{v}(\bm {x},t)=\frac {\partial}{\partial
t}\bm{\xi}
\label{Firstb}\\
\rho(\bm{x},t)=\rho(\bm{x}_{0})-
\bm{\nabla}_{0}\cdot[\rho(\bm{x}_{0})\bm{\xi}]=
\rho(\bm{x}_{0})+\delta \rho \label{Firstc}\\
\mu(\bm{x}, t) = \mu_{\text{loc}}(\bm{x}_{0}) + V_{\text{
ext}}-\frac{\partial}{\partial \rho}(\mu_{\text
{loc}})[\bm{\nabla}_{0}\cdot(\rho_{0}\bm{\xi})] \label{Firstd},
\end{eqnarray}
\end{subequations}
where $\bm{p}(\bm{x}_{0})=0$ in BEC and $\bm{\nabla}_{0}$ denotes
the partial derivative with respect to $\bm{x}_{0}$ with
$\bm{\nabla}\rightarrow
\bm{\nabla}_{0}-\bm{\nabla}_{0}\bm{\xi}\cdot\bm{\nabla}_{0}$.
Eq.~(\ref{Firstc}) was derived by substituting Eq.~(\ref{Firstb})
to Eq.~(\ref{Cont1}), then integrating over time.

These full equations of motion describe the ensemble of stable
particle trajectories from which we may the study of the perturbed
particle trajectories as we have done previously for a spherically
symmetric condensate \cite{SJHan1}. The perturbed particle orbits
are obtained by the same method as Weinberg's analysis of electron
orbits \cite{Weinberg67}.

It may also be worth of noticing, though less obviously, that
Feynman's atomic displacements with the back-flow in his intuitive
description of phonons in $^{4}$He \cite{Feynman53} are
mathematically equivalent to our approach with the Lagrangian
displacements and the phase coherence in ODLRO; we arrive at the
almost same dispersion relation for a sound (longitudinal) wave.
Hence our analysis of collective excitations in BEC is essentially
equivalent to that of atomic theory of the two-fluid model of
Feynman since both approaches study the ensemble of particle
trajectories.

In order to make a contact with a possible future experiment, we
make the following assumptions: the condensate is a long and thin
cylinder, {\it i.e.,} $L\gg b$, where $L$ is the length of the
condensate, $b$ the radius. We therefore assume $kb\ll 1$, where
$k$ is the wave number along the z-axis. We may relax the last
assumption, but with considerable amount of additional algebra. We
further assume, for simplicity, a cylindrically symmetric
condensate and $\lambda=\omega_{z}^{0}/\omega_{\perp}^{0}\ll 1$
which is consistent with $L\gg b$, and also $kb\ll 1$. These
assumptions reduce the problem effectively to a two-dimensional
problem in the cylindrical polar coordinates.

If we take the linear momentum $\bm{p}_{0}=0$ in BEC, the equation
of motion gives
\begin{equation}
\mu_{\text{loc0}}(\rho_{0}(\bm{x}_{0}),t)+V_{\text{ext}}=constant
=\mu_{\text{loc0}}[\rho_{0}(0)]=\mu_{0}, %
\label{EQS1}
\end{equation}
which shows that the peak density at the center of a trap plays a
pivotal role - an essential point, since the speed of (first)
sound $c=[4\pi\,a\rho(0)\hbar^{2}]^{1/2}/M$ must be finite at the
center and dissipates to zero giving rise to a surface energy at
the nodal surface as we shall see.

Now we are ready to derive the first-order equation of motion in
terms of $\bm{\xi}$, which describes the collective excitations
with initial-boundary conditions. To show this we linearize
Eq.~(\ref{Motion0}) using Eq.~(\ref{EQS0}), Eqs.~(\ref{First}) and
Eq.~(\ref{EQS1}) and the result is
\begin{equation}
\frac{\partial^{2}}{\partial t^{2}}\bm{\xi}
=\frac{\mu_{0}}{M}\bm{\nabla}\sigma-\omega_{0}^{2}\,[(\bm{
\xi}\cdot\bm{\nabla})\bm{r}+(\bm {r} \cdot\bm{\nabla}\bm{\xi})]
-\omega_{0}^{2}\,[\sigma\bm {r}
+\frac{1}{2}r^{2}\bm{\nabla}\sigma].%
\label{Master}
\end{equation}

Here we have taken $V_{ext}(r)=M\omega_{0}^{2}\,r^{2}/2$,
$\partial \mu_{loc}/\partial \rho=4\pi\hbar^{2}\,a/M$ \cite{FW71},
$\sigma={\bf \nabla}\cdot{\bm{\xi}}$, and $\omega_{0}$ is the
radial trap frequency, and have also dropped the subscript in
$\bm{r}_{0}$. It is also understood henceforth that
$\bm{r}=\bm{r}_{\perp}$, $\bm{\nabla}=\bm{\nabla}_{\perp}$, and
$\bm{\xi}=(\xi_{r},\xi_{\theta})e^{\imath m\theta}$.

The entire discussion of collective excitations is based on the
first-order equation Eq.~(\ref{Master}) which must be solved by
initial-boundary conditions \cite{SJHan1}. We proceed further in
two stages. First, we discuss the surface waves on the free
surface of CSIBC in a trap and show that the free surface of the
BEC in equilibrium behaves like a classical fluid. Secondly, we
solve the same equation, but remove the condition of
incompressibility to derive the dispersion relation for a
compressional wave, an ordinary sound wave (phonons) which shows
that the presence of collective excitations with excitation
energies that are linear $\omega=c\,k$ with respect to both the
speed of first sound $c=[4\pi\,a\rho\hbar^{2}]^{1/2}/M$, and the
phonon wave number $k=k_{\theta}=m/r$, consistent with the
Bogoliubov theory \cite{Bog47}.

\section{\label{sec:level1}Dispersion relation for surface waves in CSIBC}

In this section we carry out the first part of the analysis for
the collective excitations in CSIBC. The excitation of surface
waves on a free surface can be initiated by perturbations of the
external trapping potential just like gravity waves on the surface
of a fluid in a gravitational field.

For the surface waves, we may impose the following two boundary
conditions on $\bm{\xi}$ at the free surface of the condensate:
one of which is incompressibility of a fluid at the free surface
and the other of which is the condition of irrotational motion at
the free surface \cite{Landau41,Lamb45}. Since we intend to relax
the first condition later in the discussion of a compressional
wave which is our main objective in this paper, no extensive
discussion is necessary. In any case, they are mathematically
$\sigma\equiv\bm{\nabla}\cdot\bm{\xi}=0$ and
$\bm{\omega}\equiv\bm{\nabla}\times\bm{\xi}=0$. These two
conditions allow one to rewrite the first-order equation,
Eq.~(\ref{Master}) as

 \begin{equation}
 \frac{\partial^{2}}{\partial t^{2}}\bm{\xi}=
 -\omega_{0}^{2}\,\bm{\xi}-\omega_{0}^{2}\,(\bm {r}\cdot
 \bm{\nabla})\bm{\xi}.
 \label{surf1}
 \end{equation}

Also, the above two conditions imply that there exists a scalar
function $\chi$ that satisfies the Laplace equation,
$\nabla^{2}\chi=0$ and $\bm {\xi}_{\text{s}}=-\bm{\nabla}\chi$.
Here the subscript $s$ stands for the surface waves. The general
solution for $\chi$ is given in the cylindrical coordinates by

 \begin{equation}
 \chi(r, t)= [Q^{m}_{+}(t)r^{m}+Q^{m}_{-}(t)r^{-(m)}]\,
 e^{\imath m\theta}
 \label{surf2}
 \end{equation}
 where we may set $Q^{m}_{-}(t)=0$ for a solid cylindrical condensate.

It is a simple algebra, taking the gradient on Eq.~(\ref{surf2})
and substituting it into Eq.~(\ref{surf1}), to obtain the
time-dependent equation for $Q^{m}_{+}$,
\begin{equation}
\frac{d^{2}}{dt^{2}}Q^{m}_{+}+m\,\omega_{0}^{2}\,Q^{m}_{+}=0,
\end{equation}
from which one can write down the dispersion relation at once
\begin {equation}
\omega^{2}_{surf}=m\,\omega_{0}^{2}. \label{Surf}
\end {equation}

The dispersion relation for the surface waves is a function of the
radial trapping frequency (external force) $\omega_{0}$ and the
mode number m. There are two aspects to the dispersion relation.
First, the dispersion relation is independent of the the
pair-interaction potential ({\it i.e.,} internal dynamics) of a
trapped Bose gas and of the radius of the condensate. A priori,
the condition of irrotational flow is assumed to derive the
dispersion relation. Since this condition is valid for a surface
wave both in a classical fluid and in a superfluid \cite{Lamb45,
Landau63}, the dispersion relation Eq.~(\ref{Surf}) is a
manifestation of broken symmetry in a Bose system. Moreover, it is
independent of $\hbar$ in spite of the quantum ground state
density given by Eq.~(\ref{Grho}) which is a function of $\hbar$.
Second, the surface waves are driven solely by the perturbations
of an external trapping force on the surface of BEC just like a
gravity wave under the action of the force of gravity in a
gravitational field \cite{Landau63}.

Unlike the spherical wave \cite{SJHan1}, the dispersion relation
Eq.~(\ref{Surf}) has an interesting geometrical interpretation:
for $m=0$, it corresponds the sausage mode for which the fluid
column makes a radial oscillation, but this mode is absent in
CSIBC on account of $\omega=0$; for $m=1$, the hose mode, the
column makes a garden hose like motion without changing the shape
of the cross-section. For higher mode $m\gg 1$, $k_{\theta}=m/b$
corresponds a short waves along the circumference of the column
with the radius of $b$. Moreover, it is fairly simple a task to
observe the surface waves by applying the perturbation along the
waist line in a long ellipsoidal condensate. The experimental data
\cite{Onofrio00} agree well with Eq.~(\ref{Surf}).

A few final remarks will close this section. First, it is
important to note that the dispersion relation for the surface
waves in BEC [Eq.~(\ref{Surf})] also implies that Osborne's
experimental observation on the contour of a rotating He II
\cite{Osborne50} remains indeed correct, a point which appears to
be quite contrary to what one expects from Landau's two-fluid
model \cite{Landau41} and which has been extensively studied
\cite{Hall60,Feynman55} since Osborne's first observation on the
surface contour of rotating $^{4}$He \cite{Osborne50} in 1950. In
short, the free surface of a superfluid behaves like a classical
fluid under the external force; this is a peculiarly universal
behavior of a superfluid. It is also evident that Feynman's vortex
line model \cite{Feynman55} cannot explain the Osborne's
observation as emphasized by Meservey \cite{Meservey64}. Second,
Rayfield and Reif \cite{Rayfield64} observed that, apart from the
quantization of circulation, a vortex moves like a classical fluid
with an ion probe at core, which implies a break-down of
superfluidity at a vortex core \cite{Glaberson68}. This peculiar
universal classical fluid-like behavior of the free surface of
CSIBC is essentially in agreement with the breakdown of
superfluidity at a nodal surface of a vortex core. This
observation also resolves the recent controversy over the Magnus
force \cite{Thouless99}. The classical Magnus force employed by
Vinen \cite{Vinen61} in his analysis of vortex quantization is
indeed the correct one which gave the value
$\kappa=\oint\bm{v}_{s}\cdot d\bm{l}$, where
$\kappa=h/M=0.997\times 10^{-3}cm^{2}/sec$
\cite{Rayfield64,SJHan3}.

\section{\label{sec:level2}Dispersion relation for
(first) sound waves in CSIBC}

One problem that requires the introduction of the collective
excitations for its solution is that of the superfluidity of
CSIBC. Our aim here is to understand the (first) sound wave
propagation and its modified excitation spectrum in CSIBC. This
calculation gives the excitation spectrum almost identical to that
of Bogoliubov's phonon spectrum, $\omega=ck$ with
$c=[4\pi\,a\rho\hbar^{2}]^{1/2}/M$ and $k=m/r$.

It may be useful to recall Feynman's theory of phonons at this
point. In a series of papers \cite{Feynman53}, Feynman has laid
out an elegantly simple theory of collective excitations (phonons
and rotons) in $^{4}$He based on his path-integral approach to
quantum mechanics based on the exact quantum partition function.
Feynman has shown that, at temperature sufficiently below the
$\lambda$ point, the ground state wave function has a positive
amplitude for any configuration, since the ground state wave
function is symmetric and thus has no nodal surface. A density
fluctuation involving a large number of particles (or long
wavelengths) creates a back-flow by the conservation of particle
density; the compressed density in one part of the system is then
left with the rarefaction adjacently. Feynman goes on to argue
that, with the back-flow and the conservation of the momentum, the
atomic displacements do not lead to a single particle excitation,
but to extreme low-energy excitations of compressional waves with
no nodes (phonons), since a nodal surface breaks the symmetry
property of a Bose system as we have shown in Sec II of this
paper. In spite of a simple picture of sound waves in Feynman's
analysis just like Landau's two-fluid model \cite{Landau41}, the
speed of (first) sound is not given as a function of the
pair-interaction potential which he presented at the outset, but
is shown to be identical with the velocity of sound defined by the
usual macroscopic considerations \cite{Landau41}.

In contrast, the above description of our method implies that,
without any reference to a classical fluid, one should be able to
explicitly calculate the modified excitation spectrum as a
function of the speed of sound and the wave number with the
condition of irrotational motion of a superfluid. We now define
the compressibility of the fluid as $\sigma=\bm{\nabla}\cdot{\bm
{\xi}}$ in Eq.~(\ref{Firstc}). It is a mathematical description of
small-amplitude density waves (phonons) through the equation of
continuity and the phase-coherence, and is essentially equivalent
to Feynman's picture of phonons in superfluid He II. The energy
spectrum of collective excitations in the phonon regime can be
obtained from Eq.~(\ref{Master}) by taking the divergence on both
sides. With the vector identity ${\bf\nabla}\cdot[({\bf
r}\cdot{\bf\nabla}){\bf\xi}]={\bf r}\cdot{\bf
\nabla}\sigma+\sigma$, where ${\bf\nabla}={\bf\nabla}_{\perp}$ and
${\bf r}={\bf r}_{\perp}$, and the condition of irrotational
motion ${\bf\nabla}\times{\bf\xi}=0$ \cite{Landau41}, it is a
straightforward algebra to obtain the first-order equation,

\begin{equation}
\frac{\partial^{2}}{\partial t^{2}}
\sigma(r,t)=\frac{1}{2}\omega_{0}^{2}\,(\alpha^{2}-r^{2})\nabla^{2}\sigma-
3\,\omega_{0}^{2}\,r\,\frac{\partial}{\partial
r}\sigma-4\,\omega_{0}^{2}\,\sigma,
\end{equation}
where $\alpha^{2}=8\pi\,a n_{0}(0)\hbar^{2}/(M\,\omega_{0})^{2}$
and $\nabla^{2}=\nabla^{2}_{\perp}$.

Writing $\sigma({\bf r}, t)= S(t)W(r)e^{\imath m \theta}$, we
obtain the variable separated equations:
\begin{equation}
\frac{d^{2}}{dt^{2}}S(t)+\lambda_{n} S(t)=0.%
\label{Time}
\end{equation}
\begin{eqnarray}
(\alpha^{2}-r^{2})\,[\frac{1}{r}\frac{d}{dr}(r\frac{d}{dr}W_{n}(r))-
\frac{m^{2}}{r^{2}}\,W_{n}(r)]\nonumber\\
-6r\frac{d}{dr}W_{n}(r)\nonumber\\
+2(\lambda_{n}/\omega_{0}^{2}-4)\,W_{n}(r)=0,%
\label{Space}
\end{eqnarray}
where $\lambda_{n}$ is a constant of separation.

The dispersion relation is determined by the eigenvalues of the
radial equation, Eq.~(\ref{Space}). In the following we show the
eigenvalues are a function of the speed of (first) sound and the
radial trap frequency by transforming  Eq.~(\ref{Space}) to the
Sturm-Liouville problem and thereby obtaining the eigenvalues in
terms of the complete set of ortho-normal functions,
\begin{equation}
\acute{\lambda_{n}}=\int_{0}^{b}r(\alpha^{2}-r^{2})^{3}
\,[(\frac{d}{dr}W_{n})^{2}+ \frac{m^{2}}{r^{2}}\,W_{n}^{2}(r)]dr
\label{Eigen}
\end{equation}

and with

\begin{equation}
\int_{0}^{b}r(\alpha^{2}-r^{2})^{2}\,W_{m}(r)W_{n}(r)dr=\delta_{m,n}.
\label{Ortho}
\end{equation}

Since $\acute{\lambda_{n}}=2(\lambda_{n}/\omega_{0}^{2}-4)$ and
$b=[8\pi\,a\rho_{0}(0)\hbar^{2}]^{1/2}/(\omega_{0}M)$,
Eq.~(\ref{Eigen}) shows that the eigenvalues are indeed a function
of the speed of (first) sound at the center of the condensate
({\it i.e.,} $b=[\sqrt{2}/\omega_{0}]c_{ctr}$ and $c_{ctr}=[4\pi a
\rho(0)\hbar^{2}]^{1/2}/M$) for all values of the mode number $m$.
To derive Eq.~(\ref{Eigen}), we have tacitly assumed a point
source at the center of CSIBC and thus $\sigma(\bm{r})=0$ at both
$r=0$ and the free surface $r=b$. Eqs.~(\ref{Eigen}) and
(\ref{Ortho}) show that a sound wave travels from the center,
where it has the peak velocity, to the free surface since the
speed of the (first) sound wave varies as the density of CSIBC is
not uniform. The eigenvalues are also a function of the mode
number. The reason for this is of course that the mode number $m$
is a constant of the motion.

In principle the eigenvalues can be obtained from
Eq.~(\ref{Eigen}), but the integral representation is hopelessly
complicated to evaluate the eigenvalues. Nevertheless, it shows
that the eigenvalues are positive definite and they are indeed a
function of the speed of (first) sound $c_{ctr}$ for all values of
the mode number.

It is helpful to look at Eq.~(\ref{Space}) in complex plane and to
consider Eq.~(\ref{Space}) as a linear eigenvalue equation. This
can be done easily by mathematical transformation; it also serves
a very useful and important purpose to see how the sound wave
travels in CSIBC. If we define the complex variable
$W_{n}(r)=r^{\pm(m)}Z_{n}(r)$ and with $x=r^{2}/\alpha^{2}$, then
Eq.~(\ref{Space}) transforms to a well-known differential
equation,
\begin{equation}
x(1-x)\,\frac{d^{2}}{dx^{2}}Z_{n}+[c-(a+b+1)\,x]\,
\frac{d}{dx}Z_{n} -ab\,Z_{n}=0.
\label{Last}
\end{equation}
This is the Gauss's hypergeometric equation \cite{WW63} with
$c=\,\pm m+1$, $a+b=\,\pm m+3$, and
$ab=(1/4)\,[\acute{\lambda}_{n}\,\mp\,6m]$.

The advantage of this transformation is to calculate the
eigenvalues by means of a simple numerical method. Moreover, there
are none of the end-point singularity problems associated with
$W_{m}(r)$ that one encounters in the integral representation
Eq.~(\ref{Eigen}). And the various numerical methods
\cite{Jeffreys56} are available to compute the eigenvalues.

One further point must be made with regard to Eq.~(\ref{Last}).
The hypergeometric equation Eq.~(\ref{Last}) has regular singular
points at $x=0$, $x=1$ and $x=\infty$. Its solution is the
hypergeometric function which is analytic in the complex plane
with a cut from $1$ to $\infty$ along the real axis. The branch
point $x=1$ corresponds to the location of the free surface of the
fluid column of CSIBC. Parenthetically, we also note that the
analyticity of the complex variable reflects the fact that a sound
wave can travel only inside of the free surface as it must.

Since the present work critically depends on the numerical
accuracy, we may pause to examine to what extent one can take the
numerical solutions to the eigenvalue problem. One question that
arises is whether or not it is possible to see the physical
significance of the cut in the complex plane. To better understand
the analyticity, it should be emphasized that $\sigma={\bf
\nabla}\cdot{\bf \xi}$ has been interpreted as an amplitude of a
compressional wave ({\it i.e.,} collective mode) in CSIBC in which
a cylindrically symmetric sound wave travels from the the center
toward the free surface. The presence of the cut implies that
there cannot be a sound wave beyond the branch point.

The collective solution exists in a domain in which $\sigma\neq
0$. Such a solution must be analytic in the complex plane. To
check this we must be sure a solution to Eq.~(\ref{Last}) is
compatible with Eq.~(\ref{Eigen}). It is physically obvious that a
sound wave can travel in the domain $[0,1]$ in which the solution
to Eq.~(\ref{Last}) is analytic. The branch point reflects the
nodal surface at which the sound wave must dissipate to conserve
the total energy of an isolated system giving rise a surface
energy. Thus our numerical solutions in the domain $[0,1]$ will
always satisfy this physical condition; it is a unique feature of
finite systems.

Here we employ a simple numerical method \cite{Hildebrand74} to
evaluate the smallest eigenvalues corresponding to the low-energy
phonon excitations in the domain $[0,1]$. This simple numerical
method should, however, be quite satisfactory provided the
solutions are numerically stable near the branch point
\cite{Jeffreys56}.

Returning to Eq.~(\ref{Time}) and taking $S(t)=e^{i\omega t}$, we obtain the
dispersion relation as
\begin{equation}
\omega_{ph}=
\pm(\lambda_{n})^{1/2}=\pm[2\acute\lambda_{s}+4]^{1/2}\,\omega_{0}. %
\label{Phonon}
\end{equation}
where $\acute\lambda_{s}$ is the smallest eigenvalues from Eq.%
~(\ref{Last}). The ratio $\omega/\omega_{0}$ with respect to the
mode number $m$ is plotted in Fig.1.

The dispersion relation Eq.~(\ref{Phonon}) (Fig. 1.) shows that
for large $m$, $\omega/\omega_{0}$ is linear with respect to the
mode number $m$ (or wave number $k_{\theta}=m/r$). It is almost
exactly the same form of the usual dispersion relation $\omega=ck$
of a (first) sound wave \cite{Landau41, Bog47} with
$k_{\theta}=m/r$. This has the following simple interpretation:
for large values of the mode number $m$, a wave, whose wavelength
is shorter than the size of CSIBC but longer than atomic
dimensions, sees the medium as if it is an infinite medium. Thus
we recover in an essential manner the Bogoliubov dispersion
relation $\omega=c\,k$ \cite{Bog47} at long wavelengths. This
assures us that in fact the present analysis describes the
collective excitations correctly at temperature near absolutely
zero in CSIBC. For small values of the mode number, the dispersion
curve is slightly parabolic with respect to the mode number due to
geometrical effects-finite size effects. For $m=0$ mode,
$\omega/\omega_{0}=2.8153$ is a unique value in a finite space
problem. It corresponds to a uniform perturbation in polar angle
in CSIBC.

As in the case of a spherically symmetric nonuniform condensate,
the whole problem of collective excitations in CSIBC is reduced to
the question of how one measures the speed of first sound
$c_{ctr}=[4\pi a n(0)\hbar^{2}]^{1/2}/M$ (or the peak density) for
an outgoing cylindrical sound wave at various radial points in
CSIBC. This requires a precise measurement of the peak density, a
prerequisite for an experimental confirmation of BEC in a
trap.

There is one more point to be made about observing the phonon
spectrum in a spatially inhomogeneous condensate. To understand
the difficulty, let us take the simplest case $m=0$ mode. Unlike
the surface waves, it is important to perturb the medium at the
center to initiate a (first) sound wave uniformly in polar angle.
How, then, if we were to confirm BEC in CSIBC by studying the
radial propagation of the sound wave, might we try to to initiate
short waves at the center and to measure the speed of sound wave
at $r<b$. As far as we can see, this is a difficult task; it is,
nonetheless, essential to see if we can initiate the sound waves
at the center and place a probe at $r < b$, inside the nodal
surface, since the the sound waves dissipate near the nodal
surface as discussed above. The question of whether a systematic
perturbation consistent with the present theory is experimentally
feasible remains to be seen. Nevertheless, it would be necessary
to have a precise dispersion curve on the collective excitations
in CSIBC similar to that of Henshaw and Woods in He II
\cite{Henshaw61}. And it is still the only way to quantitatively
establish the realization of BEC in a trap.

There is one last point to be made about the break-down of
superfluidity in BEC at the free surface. To understand this, let
us consider what has happened to the sound wave propagating toward
the free surface. As it approaches the free surface, the particle
trajectories rapidly fluctuate due the effective quantum
mechanical potential $U_{eqmp}$ which diverges as
$M\omega_{0}^{2}\mu_{0}/D^{2}$ as
$D=[\mu_{0}-(1/2)M\omega^{2}_{0}r^{2}]\rightarrow 0$ near the free
surface \cite{Bohm52}. This fluctuation breaks up the phase
coherence of the sound wave which dissipates at the free surface
giving rise a surface energy.

The important mechanism for the symmetry breaking is that the
sound wave dissipates by the interaction with the particles in the
surface layer which is ensured by the fact that the initial state
and the final state of CSIBC always have the same energy.
Therefore, the phenomenon of the fluctuation of particles due to
$U_{eqmp}$ and the dissipation of sound waves at the surface layer
can be understood in terms of Kubo's fluctuation-dissipation
theory \cite{Kubo57}. One of our main objectives of this paper is
to understand how the symmetry of the ground state wave function
breaks down at the free surface, {\it i.e.,} a nodal surface. It
is natural to identify this underlying basic mechanism for the
symmetry breaking as \textit{a spontaneously broken symmetry} at
the free surface which accompanies phonons as Nambu-Goldstone
bosons \cite{BenLee73,Weinberg96}.

\section{\label{sec:level1}Conclusion and Discussion}

I should like to close with one final remark on the symmetry
breaking in an isolated BEC in trap. By describing the
perturbation as a Lagrangian coordinate, our analysis on the
dynamical equations is a accurate quantum mechanical treatment for
a many-body problem just as that of the quantum field theory
method by Bogoliubov for a uniform system \cite{Bog47}, provided a
correct quantum ground state function  Eq.~(\ref{Grho}) is given.
Not only does the analysis correctly predicts the low-energy
excitation spectrum for phonons, consistent with the Bogoliubov
spectrum but also shows the correct low-energy spectrum for the
surface waves in a finite inhomogeneous system confined by the
external potential.

The basis of our reasoning for the symmetry breaking at the nodal
surface follows from the observation that, in spite of the quantum
ground state given by Eq.~(\ref{Grho}), the dispersion relation
for the surface wave Eq.~(\ref{Surf}) is independent of $\hbar$
but not for the phonon spectrum of a superfluid
Eq.~(\ref{Phonon}). This is a definite proof that the gauge
symmetry of a Bose system is broken at the free surface. In the
phonon regime the Bogoliubov dispersion relation $\omega=c\,k$
holds in CSIBC with the small geometrical corrections. One also
notices from Fig.1 that there is a striking similarity between the
dispersion curves of the spherical and the cylindrical
condensates, although the angular momentum $\ell$ and the mode
number $m$ have an entirely different meaning \cite{Weinberg67}.
The salient point in our results is that the modified energy
spectrum for the phonons is a function of the speed of first
sound, $c_{ctr}=[4\pi a n (0)\hbar^{2}]^{1/2}/M$ (or the s-wave
scattering length). As Bogoliubov emphasized in his work on a
uniform Bose gas at zero temperature \cite{Bog47}, the weakly
repulsive pair-interaction potential indeed plays a crucial role
for the structure of the ground state and the superfluidity
\cite{Bog47}.

\newpage
\begin{figure}
\includegraphics{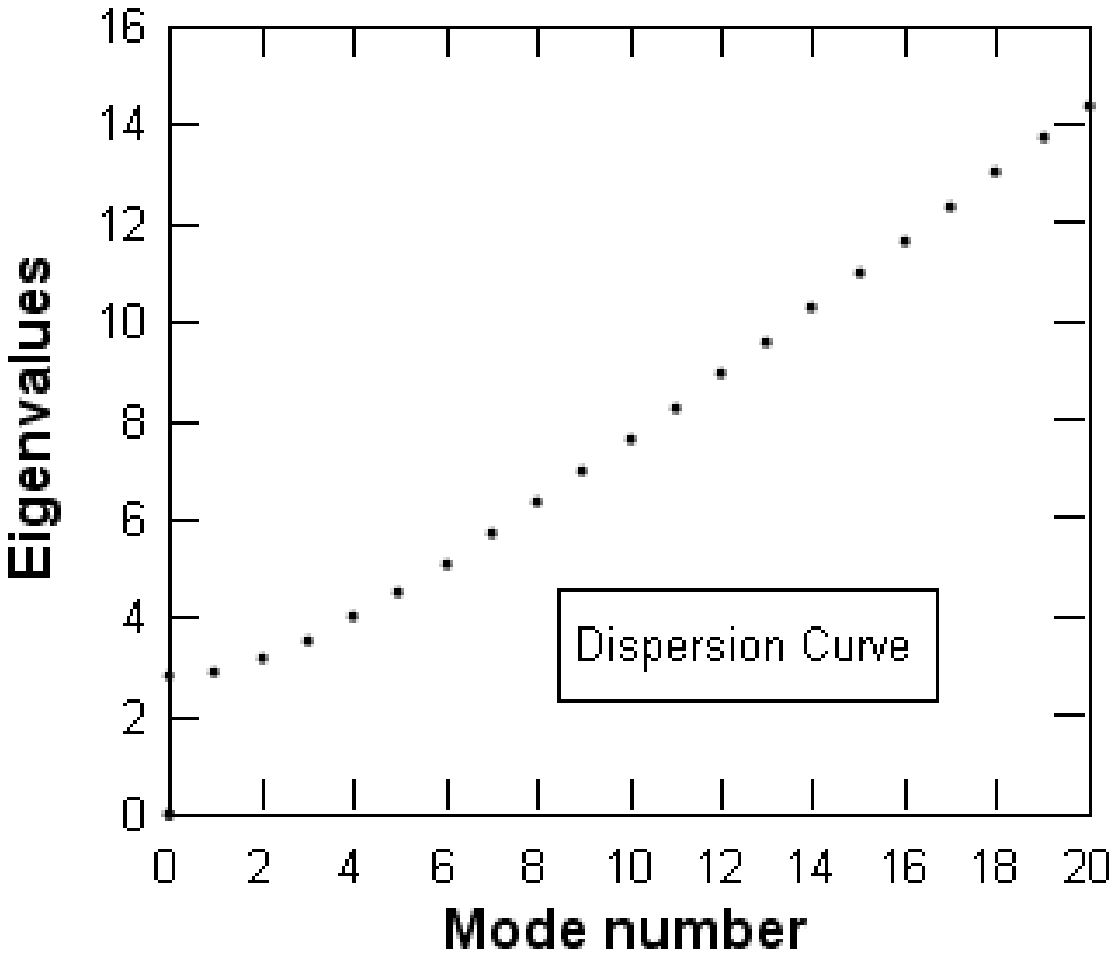}
\caption{The ratio $\omega/\omega_{0}=\lambda^{1/2}$ $
=[2\acute\lambda+4]^{1/2}$ is plotted as a function of the mode
number $m$. It shows how the energy spectrum of phonons varies
with the mode
number.}%

 \label{fig:fig1}
\end{figure}
\end{document}